\documentclass[5p]{elsarticle}
\makeatletter
\def\ps@pprintTitle{%
 \let\@oddhead\@empty
 \let\@evenhead\@empty
 \def\@oddfoot{}%
 \let\@evenfoot\@oddfoot}
\makeatother

\usepackage{csquotes}
\usepackage{graphicx,color}
\usepackage{float}
\usepackage[caption=false]{subfig}
\usepackage{amsmath}
\usepackage{amssymb}
\usepackage{multirow}
\usepackage{dsfont}
\usepackage{adjustbox}
\usepackage{pdflscape}
\usepackage{wasysym}

\begin{document}

\title{Normal high velocity solid dust impacts on tiles of tokamak-relevant temperature\vspace*{-3.0mm}}
\author{Marco De Angeli$^{a}$, Panagiotis Tolias$^{b}$,  Francisco Suzuki-Vidal$^{c}$, Dario Ripamonti$^{d}$, Tim Ringrose$^{c}$, Hugo Doyle$^{c}$, Giambattista Daminelli$^{d}$, Jay Shadbolt$^{c}$, Peter Jarvis$^{c}$, and Monica De Angeli$^{a}$}
\address{$^a$Institute for Plasma Science and Technology - CNR, via Cozzi 53, 20125 Milano, Italy\\
         $^b$Space and Plasma Physics - KTH Royal Institute of Technology, Teknikringen 31, 10044 Stockholm, Sweden\\
         $^c$First Light Fusion, Unit 10, Oxford Pioneer Park, Mead Road, Yarnton, Kidlington, Oxford, OX5 1QU, UK\\
         $^d$Institute of Condensed Matter Chemistry and Energy Technologies - CNR, via Cozzi 53, 20125 Milano, Italy\vspace*{-12.0mm}}
\begin{abstract}
\noindent Runaway electron incidence on plasma facing components triggers explosive events that are accompanied by the expulsion of fast solid debris. Subsequent dust-wall high speed impacts constitute a mechanism of wall damage and dust destruction. Empirical damage laws that can be employed for erosion estimates are based on room-temperature impact experiments. We use light-gas gun shooting systems to accelerate solid tungsten dust to near-supersonic speeds towards bulk tungsten targets that are maintained at different temperatures. This concerns targets cooled down to $-100^{\circ}$C with liquid nitrogen and targets resistively heated up to $400^{\circ}$C. Post-mortem surface analysis reveals that the three erosion regimes (plastic deformation, bonding, partial disintegration) weakly depend on the target temperature within the investigated range. It is concluded that empirical damage laws based on room-temperature measurements can be safely employed for predictions.
\end{abstract}
\maketitle

\section{Introduction}\label{sec:introduction}

\noindent Remarkable progress has been achieved in the understanding of plasma-facing component (PFC) macroscopic melt motion induced by fast transient events\,\cite{introduc1,introduc2,introduc3}, based on a combination of deliberate melting experiments\,\cite{introduc1,introduc2,introduc3,introduc4} with melt dynamics simulations\,\cite{introduc5,introduc6}. On the other hand, in spite of the fact that the potential of runaway electrons (REs) to cause excessive PFC damage due to their unique deposition profiles has been recognized from the early tokamak days\,\cite{introduc7}, a few empirical evidence and crude modelling efforts are available concerning RE-PFC interaction\,\cite{introduc8,introduc9}. Systematic observations in FTU recently revealed that unmitigated RE incidence on titanium-zirconium-molybdenum limiters drives an explosive event\,\cite{introdu01}, which ruptures the PFC and also produces fast solid dust, whose subsequent high speed mechanical impacts on adjacent tiles lead to extensive wall cratering\,\cite{introdu02}. Such an explosive scenario was very recently confirmed in controlled experiments of RE termination on graphite domes carried out in DIII-D\,\cite{introdu03}.

High velocity (HV) impacts of the newly-produced solid debris constitute a source of secondary de-localized damage induced by RE-PFC interaction, which cannot be confined into sacrificial limiters and replaceable divertor plates. Fast IR camera observations have yielded dust impact speed estimates within  $0.5-1\,$km/s in contemporary devices\,\cite{introdu01}, which could scale up to several km/s in ITER, SPARC and DEMO given the substantially larger RE currents and RE energies. The HV dust impact regime of $0.2-4\,$km/s is characterized by strong plastic deformation, partial projectile fragmentation, shallow target cratering and near-surface melting of both bodies\,\cite{introdu04}. It lies beyond the applicability range of established analytical impact models that quantify dissipation in dust-wall collisions\,\cite{introdu05,introdu06,introdu07}. This motivated systematic HV normal room temperature tungsten-on-tungsten (W-on-W) impact experiments\,\cite{introdu02}, which employed dust with diameters of some tens of microns, as empirically expected in RE-driven explosions\,\cite{introdu01,introdu03}. HV impacts were studied with light-gas guns and culminated in the extraction of reliable empirical damage laws that correlate crater dimensions with the dust speed and diameter. They were complemented by dedicated molecular dynamics (MD) simulations that provided microscopic insights on the crater formation mechanism\,\cite{introdu08,introdu09,outroref1}.

In tokamak-relevant HV impacts, the explosive nature of dust release and the harsh post-disruption environment imply that the impact angle, target temperature and projectile temperature could strongly vary. Thus, wall cratering could strongly depend on the above quantities, which casts doubt on the validity of the normal room-temperature empirical damage laws. Here, we focus on the dependence of wall cratering on the target temperature. W-on-W HV impacts have been studied in controlled experiments with two two-stage light-gas gun systems, located in First Light Fusion - Oxford (FLF)\,\cite{introdFLF} and CNR - Milano\,\cite{introdCNR}, that launch spherical nearly monodisperse W dust, with normal speeds of the order of $1-4$km/s, towards bulk W targets that are maintained at different steady state temperatures. This includes targets cooled down to $-100^{\circ}$C with liquid nitrogen (at FLF, relevant for cryogenically cooled PFCs) and targets resistively heated up to $400^{\circ}$C (at CNR, relevant for PFCs heated during the plasma plateau or the disruption evolution). The crater diameters and depths are measured by means of a scanning electron microscope and of an optical microscope. The influence of the varying target temperature on the crater dimensions and morphology is analysed. The accuracy of available room temperature correlations for the crater volume is discussed.

\section{Background}\label{sec:overview}

\noindent The \emph{high velocity (HV) range} of mechanical impacts between solid spherical micro-particles and semi-infinite bulk solid targets is roughly demarcated by $200\lesssim{v}_{\mathrm{imp}}[\mathrm{m/s}]\lesssim4000$ with the exact boundaries depending on the dust size, material temperatures, material composition and impact angle\,\cite{impactre1}. The HV impact range is squeezed between the \emph{low-to-moderate velocity range} of $v_{\mathrm{imp}}\lesssim200$ m/s, characterized by adhesive work and plastic dissipation but weak plastic deformations so that the microparticles can be assumed to retain their sphericity\,\cite{impactre2,impactre3,impactre4}, and the \emph{hypervelocity range} of ${v}_{\mathrm{imp}}\gtrsim4000\,$m/s, characterized by shockwave generation and release leading to extensive vaporization of both bodies\,\cite{impactre5,impactre6,impactre7}. Given its transitional location on the impact speed axis, the HV range is characterized by strong plastic deformation, partial dust fragmentation, impact cratering and near-surface melting\,\cite{introdu02,introdu04}.

The HV impact range can be further divided into at least three regimes\,\cite{introdu02}: the \emph{plastic deformation regime} of $200\lesssim{v}_{\mathrm{imp}}[\mathrm{m/s}]\lesssim500$ that is accompanied by severe projectile flattening, shallow target crater formation and very low rebound speeds \,\cite{impactHV1,impactHV2,impactHV3,impactHV4}, the \emph{impact bonding regime} of $500\lesssim{v}_{\mathrm{imp}}[\mathrm{m/s}]\lesssim1000$ that is accompanied by projectile sticking on the target, which is mainly achieved through metallurgical bonding and mechanical
interlocking caused by the increased local plastic deformation in the interface\,\cite{impactHV5,impactHV6,impactHV7,impactHV8}, the \emph{partial disintegration regime} of $1000\lesssim{v}_{\mathrm{imp}}[\mathrm{m/s}]\lesssim4000$ that is accompanied by material splash ejection, partial projectile fragmentation and pronounced target crater formation\,\cite{introdu02,introdu04}.

The impact bonding regime of the HV range has been extensively studied through numerical modelling\,\cite{impactHV8} and impact experiments\,\cite{impactHV0}, since it constitutes the physical basis of the cold spraying technique that has found a wide array of industrial applications\,\cite{impactHVa}. Naturally, effort has mainly focused on the estimation of the critical bonding velocity and the critical erosion velocity that delineate the impact bonding regime. Much less attention has been paid to the estimation of the crater dimensions through empirical scalings (damage laws). Such damage laws are mainly available for the hypervelocity range\,\cite{impactHVb,impactHVc} and their extrapolation to lower speeds is overly approximate. In addition, material-specific rather than general damage laws are necessary if accuracy is desired. It is noted that there is a vast literature dedicated to HV impact applications to military armour development, but it concerns non-spherical projectiles (rods) with sizes in the mm range\,\cite{impactHVd,impactHVe,impactHVf}.

In our earlier work\,\cite{introdu02}, one-stage and two-stage light-gas gun shooting systems were employed for $34$ room temperature W-on-W impact tests with three dust subpopulations ($51,\,63,\,76\,\mu$m) and normal incidence speeds spanning the entire HV range ($583-3190\,$m/s). Emphasis was put on obtaining large impact statistics. The three regimes of the HV range were identified and damage laws were obtained for the partial disintegration regime. These read
\begin{align}
D_{\mathrm{c}}&=0.0330(D_{\mathrm{d}})^{1.005}(v_{\mathrm{imp}})^{0.527}\,,\label{ourdamagediameter}\\
H_{\mathrm{c}}&=0.0000114(D_{\mathrm{d}})^{1.264}(v_{\mathrm{imp}})^{1.282}\,,\label{ourdamagedepth}
\end{align}
with $D_{\mathrm{c}}$ the crater diameter in $\mu$m, $H_{\mathrm{c}}$ the crater depth in $\mu$m, $D_{\mathrm{d}}$ the dust diameter in $\mu$m and $v_{\mathrm{imp}}$ the dust impact speed in m/s. It is noted that the crater geometry is well-approximated by a spherical cap for spherical dust, planar targets and normal impacts. Thus, damage laws can be directly used for estimates of the excavated volume\,\cite{introdu02}.

\begin{figure}
\centering
\includegraphics[width = 2.55in]{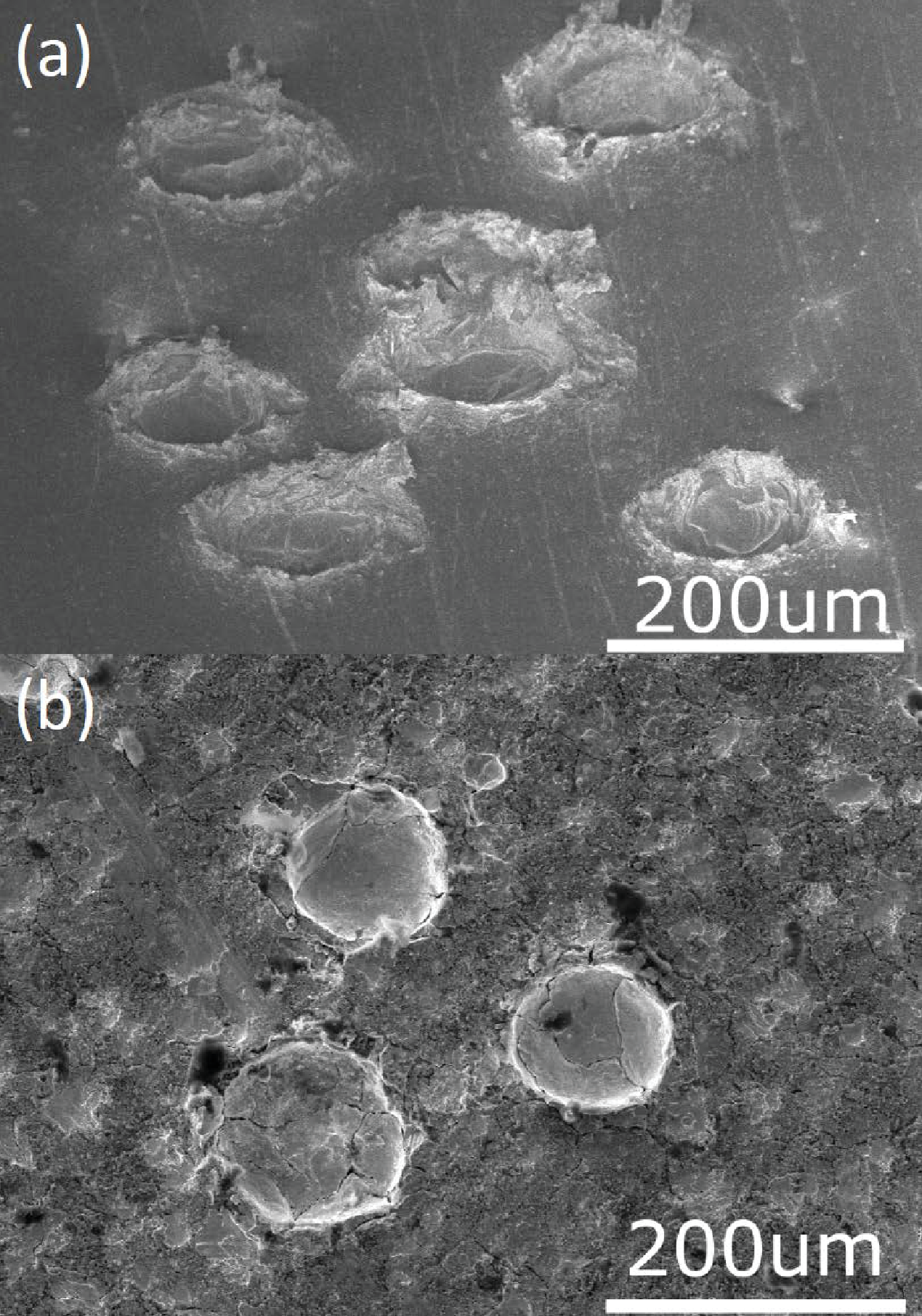}
\caption{SEM images of HV W-on-W impact craters at different target temperatures. (a) CNR 2SLGG, $u_{\mathrm{imp}}=1484\,$m/s, $T_{\mathrm{s}}=400^{\circ}\,$C ($55^\circ$ tilted). (b) FLF 2SLGG, $u_{\mathrm{imp}}=1900\,$m/s, $T_{\mathrm{s}}=-100^{\circ}\,$C.}\label{fig:multiple}\vspace*{-4.0mm}
\end{figure}

\begin{table*}
  \centering
  \caption{W-on-W impacts investigated by two different two-stage light-gas guns located at CNR and FLF. In what follows; $T_{\mathrm{s}}$ denotes the target temperature, $v_{\mathrm{imp}}$ denotes the dust impact speed, statistics refers to the number of analyzed craters, $D_{\mathrm{c}}$ is the average measured crater diameter including its standard deviation and $H_{\mathrm{c}}$ is the average measured crater depth including its standard deviation. A negative $H_{\mathrm{c}}$ designates impact sites that are elevated above the local target surface level (impact bonding regime), while a positive $H_{\mathrm{c}}$ designates impact sites that are depressed beneath the local target surface level (plastic deformation and partial disintegration regimes).}\label{Tab_collection}
\begin{tabular}{c c c c c c c }
$T_{\mathrm{s}}$  & $v_{\mathrm{imp}}$ & Statistics & $D_{\mathrm{c}}$       & $H_{\mathrm{c}}$       & Remarks                                          \\ \hline\hline
$+20^{\circ}$C    & $658.0\,$m/s       & $41$       & $91.7\pm7.7\,\mu$m     & $+10.3\pm1.1\,\mu$m    & inelastic rebound \,\,\,\,\qquad\quad\quad\quad  \\
$+200^{\circ}$C   & $627.3\,$m/s       & $23$       & $85.3\pm3.7\,\mu$m     & $+10.4\pm0.9\,\mu$m    & inelastic rebound \,\,\,\,\qquad\quad\quad\quad  \\
$+400^{\circ}$C   & $671.5\,$m/s       & $16$       & $88.2\pm5.8\,\mu$m     & $+13.4\pm1.4\,\mu$m    & $46\%$ inelastic rebound \,\,\,\,\quad\quad\quad \\
                  &                    & $19$       & $85.6\pm4.5\,\mu$m     & $-28.6\pm2.4\,\mu$m    & $54\%$ impact bonding \,\,\qquad\quad\quad       \\ \hline
$+20^{\circ}$C    & $918.0\,$m/s       & $60$       & $88.9\pm6.1\,\mu$m     & $-19.2\pm2.6\,\mu$m    & impact bonding \,\,\qquad\qquad\quad\quad        \\
$+400^{\circ}$C   & $898.5\,$m/s       & $25$       & $100.4\pm5.8\,\mu$m    & $-18.1\pm2.8\,\mu$m    & impact bonding \,\,\qquad\qquad\quad\quad        \\
$+20^{\circ}$C    & $1057.4\,$m/s      & $20$       & $95.1\pm6.1\,\mu$m     & $-12.9\pm2.1\,\mu$m    & impact bonding \,\,\qquad\qquad\quad\quad        \\
$+400^{\circ}$C   & $1019.0\,$m/s      & $25$       & $97.5\pm5.9\,\mu$m     & $-11.6\pm2.8\,\mu$m    & impact bonding \,\,\qquad\qquad\quad\quad        \\ \hline
$+20^{\circ}$C    & $1552\,$m/s        & $23$       & $108.9\pm6.3\,\mu$m    & $+19.6\pm2.6\,\mu$m    & disintegration with cracks \quad\quad            \\
$+400^{\circ}$C   & $1484\,$m/s        & $26$       & $108.8\pm7.7\,\mu$m    & $+25.3\pm3.2\,\mu$m    & disintegration with cracks \quad\quad            \\
$-100^{\circ}$C   & $1900\,$m/s        & $68$       & $109.5\pm9.5\,\mu$m    & $+38.5\pm9.9\,\mu$m    & disintegration with cracks \quad\quad            \\
$+20^{\circ}$C    & $1950\,$m/s        & $22$       & $122.1\pm4.4\,\mu$m    & $+36.3\pm2.5\,\mu$m    & disintegration with cracks \quad\quad            \\
$+200^{\circ}$C   & $2065\,$m/s        & $25$       & $126.2\pm6.9\,\mu$m    & $+41.2\pm2.1\,\mu$m    & disintegration with cracks \quad\quad            \\
$+400^{\circ}$C   & $1956\,$m/s        & $10$       & $123.6\pm6.1\,\mu$m    & $+41.1\pm2.6\,\mu$m    & disintegration with cracks \quad\quad            \\
$-100^{\circ}$C   & $2400\,$m/s        & $37$       & $135.9\pm10.1\,\mu$m   & $+65.3\pm7.4\,\mu$m    & disintegration with cracks \quad\quad            \\ \hline
$+20^{\circ}$C    & $2642\,$m/s        & $33$       & $125.8\pm5.9\,\mu$m    & $+54.2\pm7.9\,\mu$m    & disintegration without cracks                    \\
$+400^{\circ}$C   & $2739\,$m/s        & $11$       & $130.3\pm16.9\,\mu$m   & $+58.9\pm5.8\,\mu$m    & disintegration without cracks                    \\
$+20^{\circ}$C    & $3151\,$m/s        & $12$       & $135.8\pm17.4\,\mu$m   & $+64.0\pm9.3\,\mu$m    & disintegration without cracks                    \\
$+400^{\circ}$C   & $3050\,$m/s        & $43$       & $150.2\pm7.7\,\mu$m    & $+73.7\pm4.4\,\mu$m    & disintegration without cracks                    \\ \hline\hline
\end{tabular}
\end{table*}

\begin{figure*}
\centering
\includegraphics[width = 6.15in]{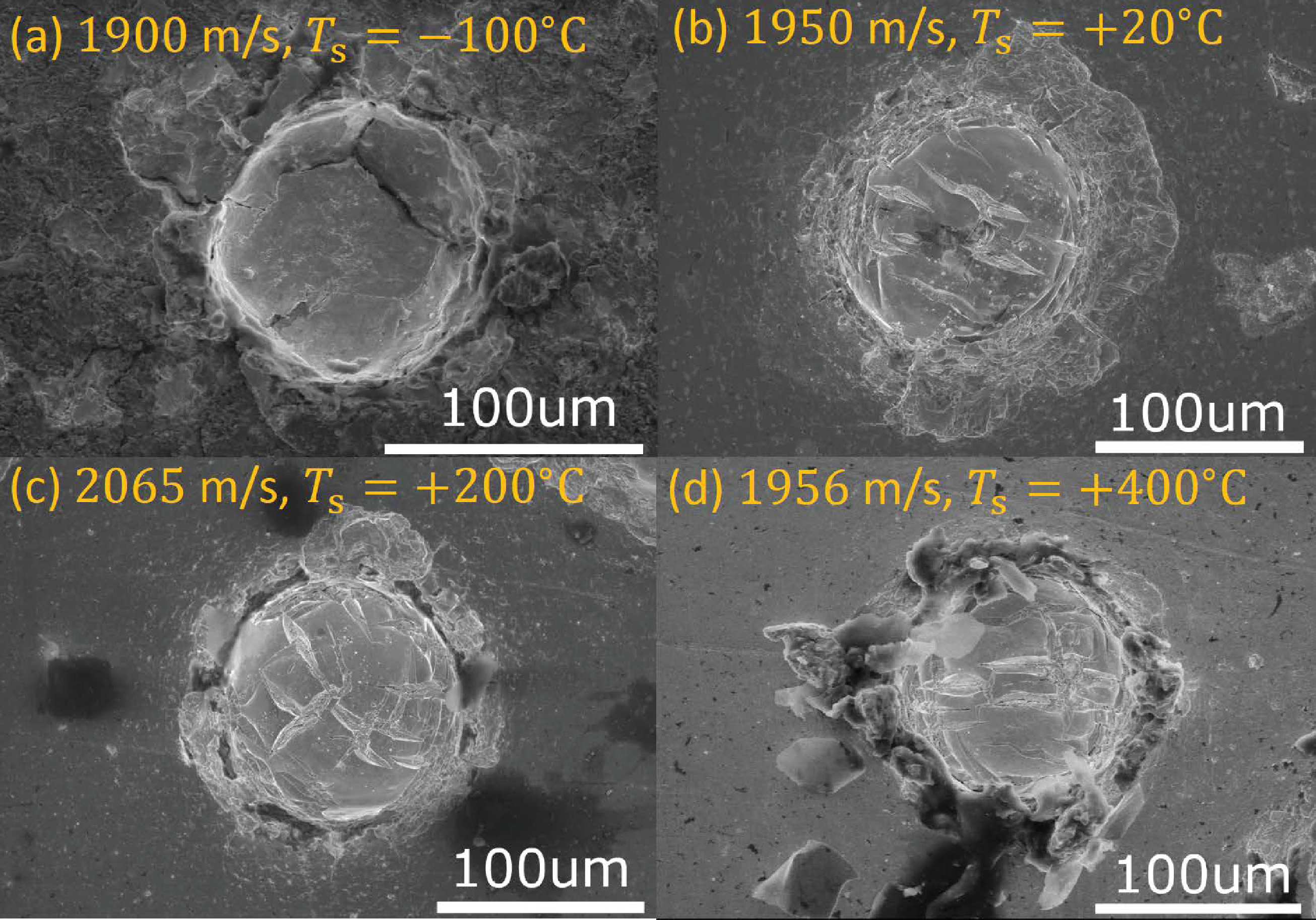}
\caption{SEM images of HV W-on-W impact craters at normal speeds of $\sim2000\,$m/s for different W target temperatures. (a) FLF 2SLGG, $u_{\mathrm{imp}}=1900\,$m/s, $T_{\mathrm{s}}=-100^{\circ}\,$C; (b) CNR 2SLGG, $u_{\mathrm{imp}}=1950\,$m/s, $T_{\mathrm{s}}=20^{\circ}\,$C; (c) CNR 2SLGG, $u_{\mathrm{imp}}=2065\,$m/s, $T_{\mathrm{s}}=200^{\circ}\,$C; (d) CNR 2SLGG, $u_{\mathrm{imp}}=1956\,$m/s, $T_{\mathrm{s}}=400^{\circ}\,$C.}\label{fig:tempdep}
\end{figure*}

\begin{figure*}
\centering
\includegraphics[width = 5.55in]{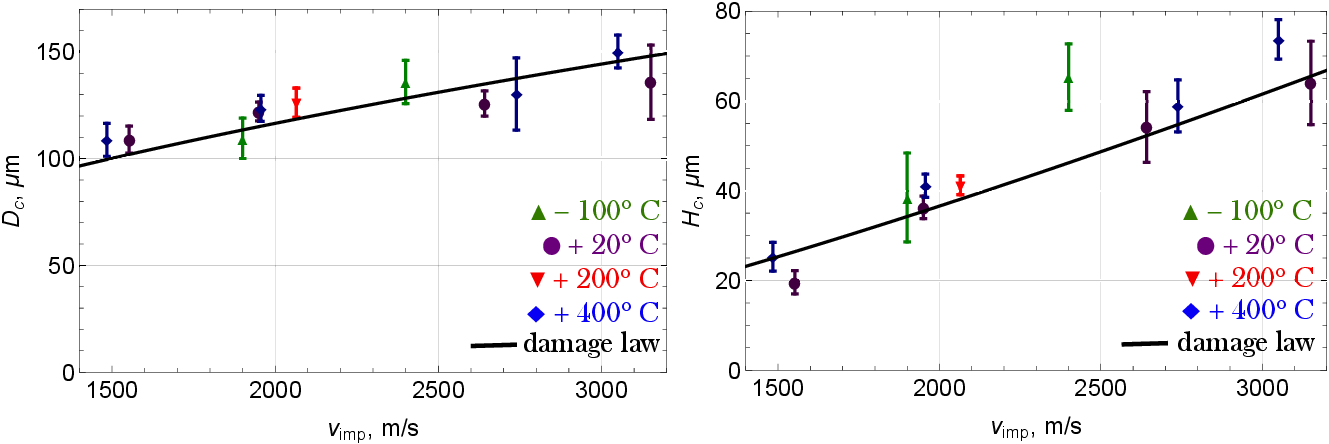}
\caption{The crater diameter (left) and crater depth (right) within the partial disintegration regime versus the normal W-on-W speed for different target temperatures. Impact tests (symbols with error bars) and the room temperature damage laws of Eqs.(\ref{ourdamagediameter},\ref{ourdamagedepth}) (solid lines).}\label{fig:comparison}\vspace*{-4.0mm}
\end{figure*}

Modelling and experimental works have focused on room temperature projectiles and targets. Nevertheless, similar to dust in tokamaks, room temperatures are rather uncommon in space applications (typically below $0^{\circ}$C) and in industrial applications (typically above $100^{\circ}$C). As far as micrometric metal projectiles are concerned, experiments have exclusively studied the impact bonding regime and specifically the dependence of the adhesion strength of the cold spray coating on the substrate temperature\,\cite{impactTE1,impactTE2}. As far as millimetric metal projectiles are concerned, only few experiments have focused on the effect of the target temperature on the crater characteristics in the high velocity and hypervelocity ranges\,\cite{impactTE5,impactTE6}. Nishida and collaborators have studied impacts of millimetric aluminum alloy spheres on thick aluminum alloy targets in the range of $0.9-3.5\,$km/s and at the target temperatures $\sim200^{\circ}$C, $\sim20^{\circ}$C and $-200^{\circ}$C with two-stage light-gas guns. They observed a significant dependence of the crater depth on the target temperatures, while the crater diameter did not exhibit a dependence\,\cite{impactTE5}. On the other hand, Ogawa and coworkers have investigated impacts of millimetric metallic spheres on thick iron meteorite and iron alloy targets in the range of $0.8-7\,$km/s and at target temperatures from $\sim20^{\circ}$C down to $-120^{\circ}$C with two-stage light-gas guns. They observed a very weak temperature dependence of all crater characteristics on the target temperature\,\cite{impactTE6}.

It is expected that varying target temperatures directly affect energy dissipation and target cratering as well as indirectly affect projectile fragmentation during HV impacts owing to (i) the strong temperature dependence of the mechanical properties relevant to plasticity and fracture, (ii) the proximity or remoteness to the liquid-solid phase transition, provided that localized melting is realized. Naturally, the effect of varying target temperature strongly depends on the material composition, which explains the seemingly conflicting conclusions of Refs.\cite{impactTE5,impactTE6}.

\section{Experimental}\label{sec:experiment}

\noindent High-sphericity low internal porosity W dust was purchased from \enquote{TEKNA Plasma Systems}. The original batch had a nominal size distribution of $45-90\,\mu$m. From this polydisperse batch, a nearly monodisperse sub-population
was meshed out using a sequence of sieves with nominal sizes of $71,\,63,\,56\,\mu$m. The mean W dust diameter is $63(\pm5)\mu$m.

W-on-W dust-wall HV impacts are studied by means of two-stage light-gas guns (2SLGGs)\,\cite{introdFLF,introdCNR}. In the second stage of both 2SLGGs, energy is converted into compressive work on the light-gas, which is utilized to accelerate a macroscopic projectile that carries dust. The required energy is released either from a high-pressure reservoir (CNR 2SLGG, gas-gas gun type) or from an ignited propellant (FLF 2SLGG, powder-gas gun type).

\emph{In the CNR 2SLGG}, as mentioned above, the first stage features a high pressure reservoir that is connected to the second stage with a fast valve, while the second stage comprises a cylinder in which the light-gas ($\mathrm{H}_2$) is fed at relatively low pressure and is compressed by a free piston. The compressed $\mathrm{H}_2$ gas rapidly expands into the launch tube simultaneously accelerating a pre-cut macroscopic projectile (sabot) which incorporates a cavity loaded with micron dust. In addition, $\mathrm{N}_2$ gas ($1$ bar) is fed between the sabot and the diaphragm located at the end of the launch tube. The shockwave generated in front of the sabot ruptures the diaphragm and splits the sabot into two pieces, releasing the loaded dust. Aiming to reduce contamination from debris due to the unavoidable HV impact of the sabot itself, the free streaming dust particles are filtered through multiple properly aligned holes prior to entering the target chamber. The target is sandwiched in a brass sample holder with a central hole. The holder could be resistively heated with a glow plug for Diesel engines and its temperature is measured on the back by a type K thermocouple.

\begin{figure}
\centering
\includegraphics[width = 3.40in]{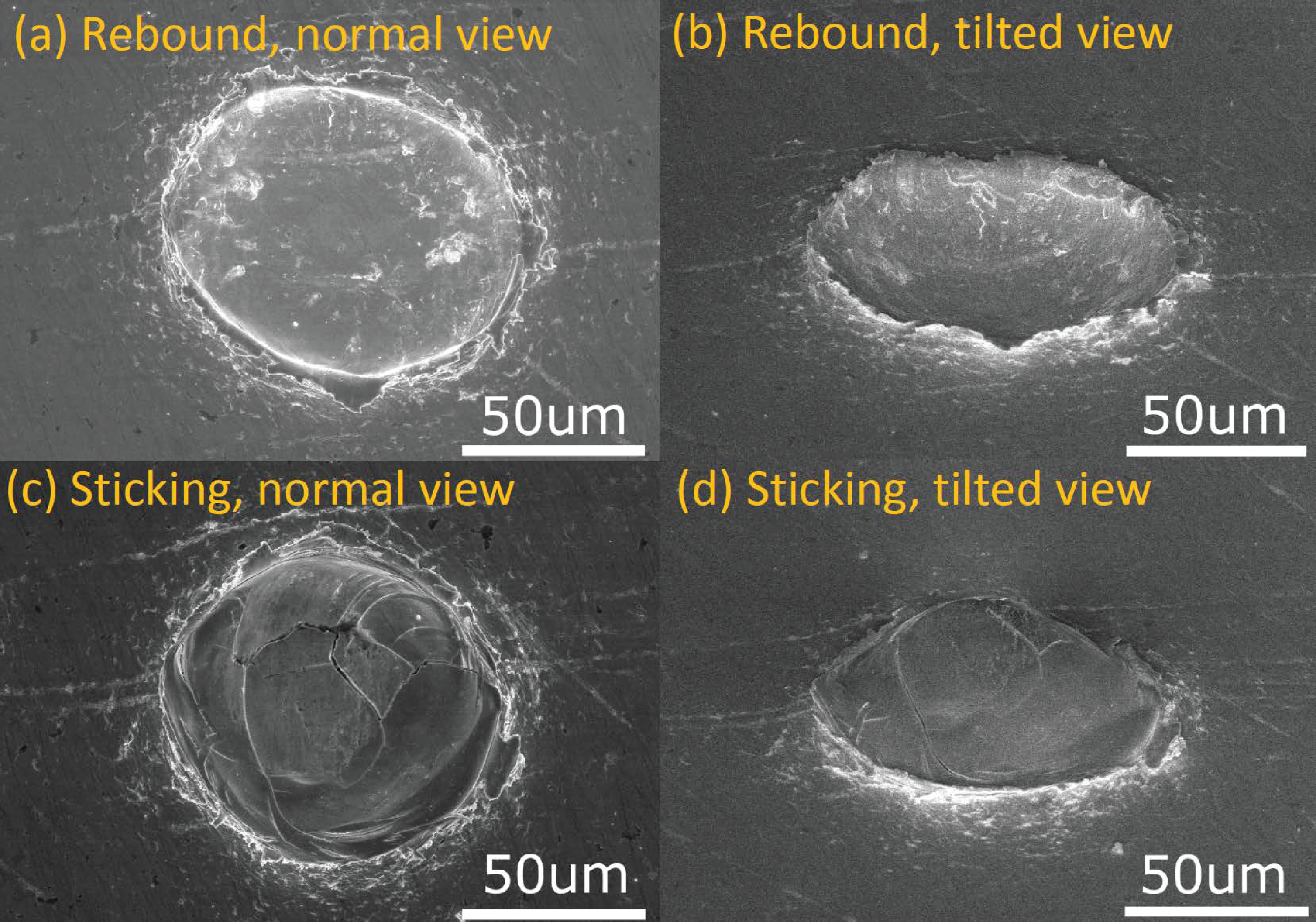}
\caption{SEM images of HV W-on-W impact craters, at a normal speed of $671.5\,$m/s and for $+400^{\circ}$C target temperature, documenting the coexistence of the plastic deformation and impact bonding regimes. Normal (left panel) and $55^{\circ}$ tilted view (right panel).}\label{fig:stickbounce}\vspace*{-4.0mm}
\end{figure}

\emph{In the FLF 2SLGG}, as aforementioned, the ignition of the powder charge launches a free piston which adiabatically compresses the propellant gas. The burst disc, separating the high pressure section from the launch tube, fails when a maximum pressure is exceeded, exposing the sabot to high pressure gas which accelerates it along the launch tube. Two part sabots are employed that are held together with plastic pins and that feature a cavity loaded with micron dust. At the end of the gun barrel, the sabot impacts on a conical sabot stripper with a hole in the middle, thus releasing the loaded dust towards the target chamber. The target is attached to a copper block that is cryogenically cooled by liquid nitrogen flow. Thermal paste is used to ensure good thermal contact. The temperature is measured on the front by a type K thermocouple that has been cross calibrated against a more accurate type T thermocouple.

\emph{In the CNR 2SLGG}, preliminary tests with large dust quantities have been carried out to confirm that the sabot and dust cloud travel at the same speeds in the range of interest. In such tests, dust cloud speeds could be measured optically through the dust transit time between two laser sheets. In the actual impact tests, a small amount of dust has been loaded to avoid overlapping craters and contamination from dust-dust impacts occurring in the proximity of the target (given the finite width of dense dust clouds), see Fig.\ref{fig:multiple}a for an example. Varying sabot ($\equiv$\,dust) speeds within the HV range of $627\lesssim{v}_{\mathrm{imp}}[\mathrm{m/s}]\lesssim3151$ have been achieved by setting different initial pressures for the first stage and controlling the gas flow resistivity between the stages. The impact speed is reproducible within $\sim50\,$m/s with the uncertainty stemming from the gas pressure precision and uncontrollable sabot friction variations in the launch tube.\,Overall, $18$ impact tests were performed at $3$ target temperatures $(20^{\circ}$C,\,$200^{\circ}$C,\,$400^{\circ}$C), out of which $16$ led to a sufficient number of craters for analysis (Table \ref{Tab_collection}).

\emph{In the FLF 2SLGG}, the emphasis was put on developing cryogenic impact test capabilities. Thus, only $2$ impact tests were performed at a $-100^{\circ}$C target temperature (Table \ref{Tab_collection}). Given the powder burn phase of the first stage, the impact speed is naturally less repeatable. Despite efforts to image the dust in flight, only the sabot speed could be measured. Therefore, it has to be assumed that the sabot speed is equal to the dust impact speed. It should also be pointed out that more debris was able to reach the target in the FLF 2SLGG tests. However, the associated craters only weakly contaminate the statistical analysis, since they can be differentiated from their much lower depths and diameters (given the much lower mass density of the sabot / stripper material compared to the W dust), see also Fig.\ref{fig:multiple}b.

All W-on-W samples were mapped by means of a Scanning Electron Microscope (SEM), at low magnification and high resolution, in order to estimate the crater diameter. The crater depth was measured with a precision optical microscope of $0.5\,\mu$m sensitivity. The instrumental uncertainty was estimated to be $\pm3\,\mu$m for the crater depth (optical) and $\pm15\%$ for the crater diameter (SEM).

\section{Results}\label{sec:results}

\noindent As seen in Table \ref{Tab_collection}, the plastic deformation, impact bonding and partial disintegration regimes were all reproduced at elevated temperatures, while only the partial disintegration regime was reproduced at cryogenic temperatures. The crater morphology (see Fig.\ref{fig:tempdep}) as well as the HV impact regime boundaries have been concluded to have a very weak dependence on the target temperature. The crater dimensions also have a very weak dependence on the target temperature within the studied $-100^{\circ}$C to $+400^{\circ}$C range.

In Fig.\ref{fig:comparison}, particular attention is paid to the partial disintegration regime that is distinguished by significant target erosion. The crater diameter room temperature damage law of Eq.(\ref{ourdamagediameter}) describes all new impact data exceptionally well, while the crater depth room temperature damage law of Eq.(\ref{ourdamagedepth}) accurately describes all new impact data with the exception of one outlier. The outlier can be attributed to the contamination of the dataset by exactly overlapping craters, which are not possible to remove from the statistical analysis without biasing the extracted mean values and their deviation. It is worth to recall that the room temperature damage laws were obtained from much higher crater statistics involving also different dust sizes\,\cite{introdu02}. However, given the observed accuracy of the room temperature damage laws regardless of target temperature, it was deemed unnecessary to increase the crater statistics and to expand to different sizes in the course of the present scoping study.

\begin{figure}
\centering
\includegraphics[width = 2.85in]{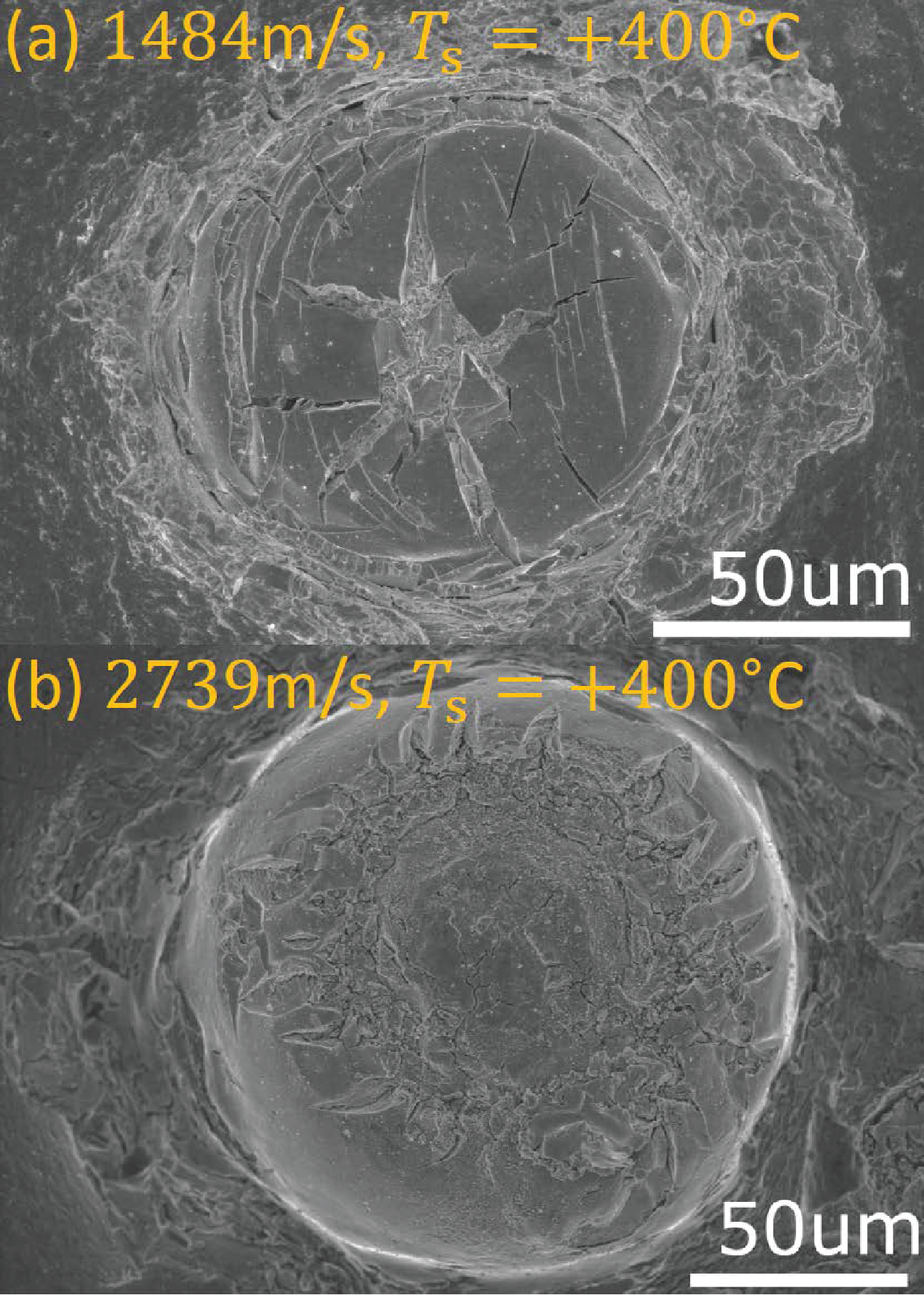}
\caption{W-on-W impact craters at the lower (a) and higher speed side (b) of the partial disintegration regime for $T_{\mathrm{s}}=400^{\circ}\,$C.}\label{fig:cracks}\vspace*{-4.0mm}
\end{figure}

A notable result concerns the overlapping of the plastic deformation and impact bonding regimes, observed at $671.5\,$m/s ($+400^{\circ}$C), see Fig.\ref{fig:stickbounce}. Under these conditions, $\sim50\%$ of the dust particles rebound and $\sim50\%$ of the dust particles stick. Thus, this impact velocity corresponds to the critical bonding velocity for the mean dust diameter of $63\,\mu$m. Being a threshold velocity, the critical bonding velocity is difficult to pinpoint experimentally. Nevertheless, in our earlier room temperature studies\,\cite{introdu02}, we had concluded that simple theoretical models\,\cite{impactHV5,impactHV6} tend to significantly underestimate it, as confirmed here. Ideally, the critical bonding velocity marks a sharp threshold between the two HV regimes, but the observed regime coexistence is justified by the fact that the nearly monodisperse W dust particles still have a narrow size distribution and the critical bonding velocity scales as $D_{\mathrm{d}}^{-0.07}$\,\cite{impactHV5,impactHV6}. Thus, dust with sizes lower than $63\,\mu$m inelastically rebounds and dust with sizes larger than $63\,\mu$m sticks. Small variations of the impact speeds around the cloud speed, small variations of the impact angle around normal incidence and small morphological variations around the nominal sphericity level should contribute to the coexistence but to a lesser degree.

Another important result concerns the crater morphology in the partial disintegration regime. It has been consistently observed that, regardless of the target temperature, the crater valley features few deep thick cracks for impact speeds roughly within $1500-2500\,$m/s that are substituted by a densely connected rugged pattern without cracks for impact speeds roughly above $2500\,$m/s, see Fig.\ref{fig:cracks}. This transition is correlated with a transition of the crater valley surface from concave to convex. This led us to conjecture that the thick cracks are formed on fragments of the partially disintegrated projectile that remain adhered to the target and that the rough pattern is primarily formed on the target itself. Therefore, the transition from the impact bonding to the partial disintegration regime is gradual, with some projectile fragments remaining adhered to the target at the low speed side of the regime. In order to confirm our hypothesis, we also performed room temperature HV impact tests within $1000-2000\,$m/s with W dust on bulk Mo targets. Backscattered electron imaging and energy dispersive X-ray analysis confirmed that projectile fragments were adhered only at the lower tested speeds.

\section{Conclusion}\label{sec:conclusions}

\noindent The high velocity normal impacts of spherical room temperature micrometric W dust on bulk W targets of varying temperature were investigated by means of two-stage light-gas guns. For target temperatures from $-100^{\circ}$C to $400^{\circ}$C, regardless of the impact speed, it is concluded that the crater morphology and dimensions have a very weak dependence on the target temperature. Hence, established empirical damage laws for the crater depth and diameter, based on room-temperature measurements, can be safely employed for erosion estimates. It is emphasized that this is valid for the studied range and does not apply for any target temperature. Note that our range includes the W ductile-to-brittle transition temperature ($\sim350^{\circ}$C)\,\cite{outroref2}, lies right below the W knee temperature ($\sim450^{\circ}$C) at which screw dislocations have similar mobility to edge dislocations\,\cite{outroref3} and lies well below the W recrystallization temperature ($\sim1100^{\circ}$C)\,\cite{outroref4}. Our results are consistent with the predictions of a recent molecular dynamics investigation for nanometric W dust which revealed that target temperatures exceeding $700^{\circ}\,$C are necessary to appreciably affect wall cratering\,\cite{outroref1}. Future work will focus on experimental investigations of the effect of oblique impact angles on the crater morphology and dimensions.

\section*{Acknowledgments}

\noindent The work has been performed within the framework of the EUROfusion Consortium,\,funded by the European Union via the Euratom Research and Training Programme (Grant Agreement No\,101052200 - EUROfusion). Views and opinions expressed are however those of the authors only and do not necessarily reflect those of the European Union or European Commission.\,Neither the European Union nor the European Commission can be held responsible for them.

\end{document}